\documentclass[11pt]{article}
\usepackage{fleqn}
\usepackage[dvips]{graphicx}
\begin{document}

\title
{\bf 
Is the nuclear force short range ?
}
\vspace{20mm}
\author{Tetsuo Sawada \\
{\small \em Institute of Quantum Science, Nihon University, 
Tokyo, Japan 101-8308}
\thanks{Associate member of IQS for research. \ \ \  
  e-mail address: t-sawada@fureai.or.jp}}
\date{}

\maketitle
\vspace{70mm}
\begin{flushleft}
{\large\bf Abstract}
\end{flushleft}

 Possible strong Van der Waals force is searched in the 
nuclear interaction, whose existence is expected when the 
fundamental interaction between quarks is strong or 
superstrong Coulombic type.     The relation between 
the type of the extra singularity at $t=0$ of 
the scattering amplitude and the power of the asymptotic 
behavior of the long range potential is derived. The square 
root cusp in the once subtracted S-wave amplitude,
 whose existence is expected when the Van der 
Waals force of the London type is acting, is observed 
when we analyze the high precision phase shift 
data of the low energy proton-proton scattering. 

\newpage

\setcounter{tocdepth}{4}

\section { Van der Waals force in the composite model of hadron }

   Before we accepted the composite model of hadron in 1960's, 
the nucleons were the elementary particles and the nuclear force 
was essentially the short range force arising from the exchanges 
of a pion and a set of pions and heavier particles.    However the 
short range nature of the nuclear force is not necessarily correct 
in composite hadron, especially when the fundamental force which 
combines the quarks is Coulombic.     It is strange that 
even after the acceptance of the composite model, the nuclear force    
is still regarded as short range with argument that,  
when the momentum transfer is not very large and one 
does not explore the inside of the composite system, we can regard 
the nucleon as elementary particle with good approximation. 
However this argument is not true, especially when the fundamental 
force is Coulombic.  In such a case, as we 
shall see later, the Van der Waals interaction between hadrons 
must always appear.  It is instructive to review the mechanism of 
the appearance of the Van der Waals potential between the neutral 
'atoms' and to estimate the strength $C$ of the potential in the 
asymptotic region 
\begin{equation}
  V(R) \sim - \frac{C}{R^6}  \qquad in \quad R \gg a \quad ,
\end{equation}  
where $a$ is the radius of the composite particle.

    Let us consider a simple system in which two atoms are in the 
ground states of $L=0$, and they are fixed with separation $R$.  
Because 
of the quantum fluctuation, atoms can also stay in the excited $L=1$ 
state, whose duration is $\Delta t \approx \hbar /\Delta E_{1}$ by the 
uncertainty principle where $\Delta E_{1}$ is the excitation energy. 
Suppose that atom 1 is in the excited P-state and therefore has the 
dipole moment $\vec{d}_{1}$.   The dipole must produce the field 
$\vec{G}$ at the location of atom 2, which decreases as $ R^{-3}$ for 
large $R$.  The field $\vec{G}$ induces a dipole $\vec{d}_{2}$ of the 
atom 2, which is $\alpha_{2} \vec{G}$ where $\alpha_{2}$ is the 
polarizability of the atom 2.   
The induced dipole $\vec{d}_{2}$ 
produces in turn a field $\vec{G'}$ at the location of atom 1, and 
therefore the energy of the system changes by $- \vec{{d}_{1}} \cdot 
\vec{G'}$ which decreases as $R^{-6}$ for large separation $R$. 
This is the mechanism of the appearance of the attractive Van der 
Waals potential of the London type.   
In particular for the case of the ordinary 
atoms, since the electric charge is the source of the constructive 
Coulomb field, the dipole $\vec{d}$ is the electric dipole and the 
field $\vec{G}$ is the electric field.  However the mechanism to 
induce the Van der Waals force is more general, and any charge, 
which produce the fundamental Coulombic force, can do the same job.  
For example, the color charge in QCD or the magnetic charge in the 
dyon model\cite{dyon} 
is responsible for the appearance of the Van der Waals 
interaction between hadrons.

\begin{figure}
   \begin{center}
      \begin{picture}(350,90)
          \put(30,30){\circle{20}}
          \put(170,30){\circle{20}}
          \put(60,60){\circle*{6}}
          \put(170,60){\circle*{6}}
          \put(170,40){\vector(0,2){17}}
          \put(37,37){\vector(1,1){21}}
          \thicklines
          \put(40,30){\vector(1,0){120}}
\put(5,20){A}
\put(185,20){B}
\put(55,70){a}
\put(170,70){b}
\put(100,10){$\vec{R}$}
\put(35,50){$\vec{r}_{1}$}
\put(175,45){$\vec{r}_{2}$}
\put(215,50){\small Fig.1.  System of atom 1 (a-A) and }
\put(242,35){\small atom 2 (b-B) separated by $\vec{R}$.}

      \end{picture}
   \end{center}

\end{figure}

     In order to estimate the strength $C$ of the Van der Waals 
potential $V(R) \sim -C/R^6$, let us consider the simplest case in 
which atom 1 and atom 2 are respectively composed of two particles 
of opposite sign.  As shown in fig.1, $\vec{r}_{i}$ is the relative 
coordinate of atom i.   The interaction Hamiltonian $H'$ of the 
system of atom 1 and atom 2 is 
\begin{eqnarray}
H' &=& ^{*} e^2 \{\frac{1}{R}+\frac{1}{|\vec{R}+\vec{r_2}-\vec{r_1}|}-
\frac{1}{|\vec{R}-\vec{r_1}|}-\frac{1}{|\vec{R}+\vec{r_2}|} \} 
\nonumber \\
&=& \frac{^{*} e^2}{R^3} (x_{1} x_{2} +y_{1} y_{2}-2 z_{1} z_{2})+
o(\frac{1}{R^5})
\end{eqnarray}  
, where $^{*} e^2$ is the "fine structure constant" of the relevant 
charge which produces the fundamental Coulombic field.  The energy 
shift $V(R)$ of the two-atom system is 
\begin{equation}
V(R)= - \sum_{n \neq 0} \frac{(H')_{0n}(H')_{n0}}{E_{n}-E_{0}} + 
\cdots .
\end{equation}  
If we see that the numerator of each term of the summation of Eq.(3) 
is positive 
definite and the energy denominators are not less than the first 
excitation energy $\Delta E_{1}=E_{1}-E_{0}$, we can obtain the 
lower and the upper bounds of the coefficient $C$ of the Van der 
Waals potential by retaining only the first term and by replacing 
all the denominators by $\Delta E_{1}$ respectively, and which is 
\begin{equation}
 \frac{|<1|(R^3 H')|0>|^2}{\Delta E_{1}} \leq C \leq 
  \frac{<0| (R^3 H')^2 |0>}{\Delta E_{1}}  \quad ,
\end{equation}  
in the estimation of the upper bound the closure relation is used.
If we substitute Eq.(1-2), the upper and the lower bounds become 
\begin{equation}
 \frac{2}{3}\frac{^{*}e^4}{(\Delta E_{1})} \hat{r}_{1}^2 \hat{r}_{2}^2
  \leq C \leq 
 \frac{2}{3}\frac{^{*}e^4}{(\Delta E_{1})} \bar{r_1}^2 \bar{r_2}^2
 \quad ,
\end{equation}  
where $\bar{r_{i}}^2$ is the ordinary mean square radius of the ground 
state of atom i.   On the other hand, $\hat{r}$ is the radius, 
which relates to the transition amplitude, defined by  
\begin{equation}
\hat{r}^2 = 3 |\tilde{z}|^2 \qquad with \qquad \tilde{z}=
\int d^{3} r \psi^{*}_{1,1,0} (r cos \theta) \psi_{0,0,0} \quad .
\end{equation}  
Now we can estimate the order of magnitude of the strength $C$ of 
the Van der Waals potential, in which $\Delta E_{1} = 2$ and 
the radius is $1/2 \sim 1/6$ in the unit of the pion mass or the pion 
Compton wave length.   The largest value of the radius 1/2 comes 
from the charge radius of the nucleon, whereas the smallest radius 
 1/6 is obtained from the distance of the repulsive core 1/3 of the 
 nucleon-nucleon interaction.   It is interesting that in QCD, in 
which $^{*}e^2 \sim 0.3$, the strength $C$ becomes 0.003 even if  
the radius is 1/2.   On the other hand in the magnetic monopole 
model of hadron,\cite{dyon} 
in which $^{*}e^2 \approx 137.0/4$, $C$ becomes 
order of 1 when the radius is 1/4$\sim$1/5, 
and in this case the Van der Waals interaction 
can compete with the nuclear force arising from the meson exchanges.

\section{ Can the nuclear potential be derived from the meson 
theory ? }

  In 1950's  and 60's, the nuclear potential had been studied 
expensively, in which Taketani's idea played an important roll.  
In his idea the nuclear force is divided into three parts according 
to the range.  
The outermost region is described by the one-pion exchange, whereas 
the innermost region is treated by phenomenological way 
and the hard core potential is used in $\mu R < x_{c}$. 
On the other hand in the middle region, the two-pion exchange is 
responsible for the nuclear potential, and which is supposed to be 
calculated from the interaction Lagrangian 
$L_{\pi,N}$ of the meson theory.  
In figure 2 the central potential $^{1}V_{C}^{+}(R)$ 
of the singlet even  state is plotted against $R$. 
The curve OPE is the one-pion exchange contribution.  
The curve TMO is the OPE plus the two-pion exchange contribution 
which is computed by the standard perturbation (static), whereas 
HM is the same quantity with full recoil considered.\cite{tmohm}  
 \vspace{5mm}

\setcounter{figure}{1}
 \begin{figure}[htbp]
 
\includegraphics[width=.93\textwidth,height=8.0cm]{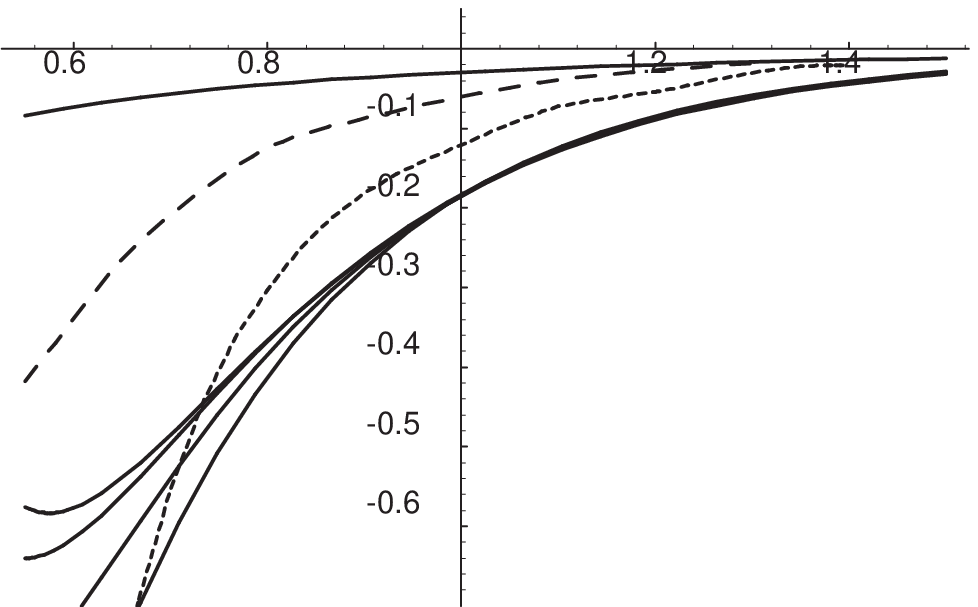}

\vskip-83mm
\hskip67mm
{\large $^{1}V^{+}_{C}(R)$ }

\vskip-2.5mm
\hskip127.mm
{\Large $R$}

\vskip7.mm
\hskip-10mm
{\large $OPE$}

\vskip15.mm
\hskip6mm
{\large $HM$}

\vskip0.mm
\hskip22mm
{\large $TMO$}

\vskip21.mm
\hskip-6mm
{\large $(d)$}

\vskip3.mm
\hskip-5mm
{\large $(c)$}

\vskip6.mm
\hskip8mm
{\large $(b)$  $(a)$}

\vskip5mm

\caption{{\footnotesize 
 Potential $^{1}V^{+}_{C}(R)$ is plotted against R in the unit of 
 the neutral pion mass and its Compton wave length.  The curves 
 a(0.30), b(0.25), c(0.20) and d($0 \sim 0.15$) are the one-boson 
 exchange potentials of Eq.(2-1) fitted to the experimental data 
 with various hard core radii, which are given in the brackets. 
 The curve OPE is the one-pion exchange potential.
 The curve TMO is the OPE plus the two-pion exchange potential which 
 is computed by the perturbation theory (static), whereas the curve 
 HM is the same type potential with full recoils are involved.
 Some attractive potential seems to be missing from the theoretical 
 curves (TMO and HM) in the region around $R \sim 1$.
}}

\end{figure}

On the other hand, curves a(0.30), b(0.25), c(0.20) and 
d($0 \sim 0.15$) are the one-boson exchange potentials: 
\begin{equation}
^{1}V_{C}^{+}=\left\{  \begin{array}{ll}
  \mu \{-(\frac{f^2}{4 \pi}) \frac{e^{-x}}{x}+A \frac{e^{-4 x}}{x}+
   B \frac{e^{-5.5 x}}{x} \} & \mbox{for $x \geq x_{c}$} \\
   \infty & \mbox{for $x < x_{c}$}
                       \end{array}
               \right.
 \end{equation}  
in which $x=\mu R$ and the parameters $A$ and $B$ are determined by 
fitting to the phase shift data, and whose values are listed in the 
table below.  The four curves differ by the 
hard core radii $x_{c}$, and whose values are written in the bracket. 

\vspace{2mm}

\begin{tabular}{|cccc|}
\hline
$\qquad curve \qquad$ & $\qquad x_{c} \qquad$ &  $\qquad A \qquad$   & 
$\qquad B \qquad$  \\ \hline
 (a)  &   0.30  &  -12.1 & 16.2 \\
 (b)  &   0.25  &  -13.6 & 23.0 \\
 (c)  &   0.20  &  -14.2 & 26.1 \\
 (d)  & 0 $\sim$ 0.15  &  -14.8 & 28.2 \\
\hline

\end{tabular}

\vspace{3mm}

It is evident that large attractive potential is missing from 
the TMO and HM potentials, and therefore these theoretical potentials 
can not reproduce the phase shift.  In order to settle the 
discrepancy, it is customary to add the term of the fictitious  
$\sigma$-meson exchange, in which the $\sigma$-meson mass 
$m_{\sigma}$ 
and the $\sigma$-N coupling constant are free parameters. 
 However in this paper we shall examine 
  whether we can attribute the 
missing attractive force to the strong Van der Waals interaction 
mentioned in the previous section.

\section{Extra singularity of the amplitude characteristic to the 
long range interaction }

   In order to identify the long range interaction in the nuclear 
force, it is important to find the behavior of the scattering 
amplitude $A(s,t)$ characteristic to the existence of the long 
range interaction.  When the nuclear force arises from the meson 
exchanges, $A(s,t)$ has a pole at $t=1$ and the continuous spectrum 
 $A_{t}(s,t)$ starting from $t=4 $ and so on.  
  What is important is that 
 in such a case $A(s,t)$ is regular at $t=0$.   On the other hand, 
 when the long range force is acting, an extra singularity appears at 
 $t=0$ and the spectral function becomes non-zero from $t=0$.  Since 
$t=0$ is the end point of the physical region $-4 \nu \leq t \leq 0$, 
where experimental data are available, 
we can in priciple determine the type and the magnitude of the 
singularity at $t=0$, when the precise data are available in the 
small neighborhood of $t=0$.   
 The threshold behavior of the spectral function $A_{t}$ is specified  
by the power $\gamma$ and the coefficient $C'$, which are defined by 
\begin{equation}
A_{t}(s,t) = \pi C' t^{\gamma} (1+o(t) )  \quad .
\end{equation} 
These parameters relate to the parameters $\alpha$ and $C$ 
of the potential $V(R)$ which are defined by $V(R) \sim - C/R^{\alpha}$
in the asymptotic region.   The relations are 
\begin{equation}
\alpha= 2 \gamma +3  \qquad and \qquad C=\frac{2C'}{m^2}
\Gamma(2 \gamma +2) \quad ,
\end{equation} 
where $m$ is the reduced mass of the potential 
scattering.\cite{present}

      It is not difficult to derive the relations, if we remember 
that the potential $V(R)$ can be represented as the superposition of 
the Yukawa potentials:
\begin{equation}
V(R)=- \frac{1}{\pi m^2} \frac{1}{R} \int_{0}^{\infty} 
w(t') e^{-R \sqrt{t'}} dt'  \quad .
\end{equation} 
Since the range of the integration is from 0 to infinity, this is 
nothing but the Laplace transformation of $R V(R)$.   What is 
important is that the weight function $w(t)$ is the imaginary part 
of the first Born amplitude $A^{(B)}(s,t)$.  we can see this simply  
by computing the Fourier transformation of both sides of Eq.(10) and 
which leads to a dispersion relation of the form 
 \begin{equation}
A^{(B)}(s,t)=\frac{1}{\pi} \int_{0}^{\infty} \frac{w(t')}{t'-t} dt'
 \quad .
\end{equation}  
In particular if the weight function is 
$w(t')=\pi C' t'^{\gamma}$, 
the potential of Eq.(10) becomes 
\begin{equation}
V(R)=-\frac{2C'}{m^2} \Gamma(2 \gamma +2) \frac{1}{R^{2 \gamma+3}}
 \quad .
\end{equation}  
 Since the change of the weight function at finite $t'$ does not 
alter the tail of the potential, we obtain the relations between 
the powers $\alpha$ and $\gamma$ and also between the coefficients 
$C$ and $C'$ given in Eq.(9). 
 
    Therefore the extra singularity of the scattering amplitude 
$A(s,t)$ at $t=0$ is 
\begin{equation}
A(s,t)=- \frac{\pi}{\sin \pi \gamma} C' (-t)^{\gamma} + \cdots 
 \quad ,
\end{equation}  
where dots mean the background regular function and can be represented 
by a polynomial of $t$.   When $\gamma$ is an integer $n$, we must 
take the limit $\gamma \rightarrow n$, and the 
singular term becomes $C' t^{n} \log (-t)$.   Since $t=-2 \nu (1-z)$, 
there are two ways to observe the extra singularity at $t=0$.
One is to see the angular distribution for fixed $\nu$ and to observe 
the singular behavior $(1-z)^{\gamma}$ at $z=1$.  Second is to make 
the partial wave projection and to observe the singularity 
$\nu^{\gamma}$ at the threshold in the partial wave amplitude 
$a_{\ell}(\nu)$.  In our normalization, $a_{\ell}(\nu)$ 
relate to the phase shifts in the elastic region by   
\begin{equation}
a_{\ell}(\nu)=\frac{\sqrt{s}}{2 \sqrt{\nu}} e^{i \delta_{\ell}(\nu)} 
 \quad ,
\end{equation} 
or in terms of the effective range function $X_{\ell}(\nu)$ 
\begin{equation}
a_{\ell}(\nu)=\frac{1}{X_{\ell}(\nu)-i \frac{2 \sqrt{\nu}} {\sqrt{s}}} 
\quad where \quad X_{\ell}(\nu)=\frac{2 \sqrt{\nu}} {\sqrt{s}} 
\cot \delta_{\ell}(\nu)
 \quad .
\end{equation} 
In order to make the extra singularity more visible in the search 
we shall use the once subtracted partial wave amplitudes which
are defined by 
\begin{equation}
a_{0}^{once}(\nu)=\frac{a_{0}(\nu)-a_{0}(0)}{\nu} \quad and \quad 
a_{\ell}^{once}(\nu)=\frac{a_{\ell}(\nu)}{\nu} \quad 
 for  \quad \ell \geq 1
 \quad .
\end{equation} 
In $a_{\ell}^{once}(\nu)$, the singularity is changed to $C''_{\ell}
\nu^{\gamma-1}$.  Therefore the extra singularity of the Van der Waals 
interaction of the London type must appear as  the square root cusp 
$C''_{\ell} \nu^{1/2}$ at $\nu=0$ . 
Moreover since the Van der Waals force is  attractive, namely $C'>0$, 
the cusp must point downward. 

    In order to observe the extra singularity as clearly as possible, 
it is necessary to remove the nearby singularities and prepare 
the domain of analyticity of the back ground function 
as wide as possible.   First of all we must 
remove the unitarity cut, whose spectral function is 
Im $a_{\ell}(\nu)=(\sqrt{s}/2 \sqrt{\nu}) \sin ^{2} \delta_{\ell}
(\nu)$, or 
in terms of the effective range function $X_{\ell}(\nu)$ 
\begin{equation}
\mbox{Im }a_{\ell}(\nu)=\frac{2 \sqrt{\nu}}{\sqrt{s}} \frac{1}
{X_{\ell}(\nu)^2 +4 \nu /s} \quad with \quad X_{\ell}(\nu)=
\frac{2 \sqrt{\nu}}{\sqrt{s}} \cot \delta_{\ell}(\nu) \quad .
\end{equation}  
This is done by introducing a function 
\begin{equation}
K_{\ell}^{once}(\nu)= a_{\ell}^{once}(\nu)- \frac{1}{\pi} \int_{0}^{U}
d \nu' \frac{\mbox{Im} a_{\ell}(\nu')}{\sqrt{\nu'} (\nu'-\nu)} \quad ,
\end{equation} 
in which the domain of the integration $[0,U]$ may be chosen 
as the region where the reasonably accurate phase shift data are 
available.

The amplitude $K_{\ell}(\nu)$ is sometimes called Kantor amplitude, 
which is free from the singularity at $\nu=0$ when all the forces 
are short range.   
Concerning the left hand cuts, the nearest one is the cut 
of the one-pion exchange which starts at $\nu=(-1/4) $.  Such a cut is 
removed in $(K_{\ell}^{once}(\nu)-a_{\ell}^{once,1 \pi}(\nu))$, where 
\begin{equation}
a_{\ell}^{once,1 \pi}(\nu)=(\frac{1}{4}\frac{g^2}{4 \pi})
\frac{1}{\nu}( \frac{1}{2 \nu} Q_{\ell}(1+\frac{1}{2 \nu}) - 
\delta_{\ell,0}) \quad .
\end{equation}  
The next nearest left hand singularity appears at $\nu=-1$ which 
is the threshold of the two-pion exchange.   Since the threshold 
behavior of the spectrum is $C''_{2 \pi}(-1-\nu)^{3/2} \theta
(-1-\nu)$, the distortion of the background function  arising from 
the two-pion exchange in the small neighborhood of $\nu=0$ 
is small.  Therefore we can observe the extra singularity at 
$\nu=0$ unless its coefficient is not very small.

\section { Observation of the cusp at $\nu=0$ in the 
once subtracted S-wave amplitude }

     In order to distinguish the effect of the long range force from 
that of the two-pion exchange, we need accurate data.  It is well 
known that the accuracy of the measurements of the low energy 
proton-proton scattering is highest in the hadron physics.   
Therefore we shall analyze the once subtracted S-wave amplitude of 
the p-p scattering $a_{0}^{once}(\nu)$, which is spin-singlet. 
Because of the electromagnetic interaction between protons, some 
modifications are necessary when we construct the once sutracted 
Kantor amplitude $K_{0}^{once}(\nu)$ given in Eq.(18). 
 Contrary to the phase shift 
$\delta_{0}(\nu)$ of the scattering by purely short range potential, 
in which the incoming and the outgoing waves are the partial wave 
projection of the plane wave, we must use the electromagnetically 
distorted wave as the initial and the final states.  The information 
of the scattering of $\ell=0$ state is summarized in the phase shift 
$\delta_{0}^{C}(\nu)$, $\delta_{0}^{E}(\nu)$ or $\delta_{0}^{EM}(\nu)$.
The small differences of these phase shifts are due to what portion 
of the electromagnetic interaction is involved when we compute the 
distorted incoming and outgoing waves.   The potentials which 
distort the initial and the final waves are: (C) pure Coulomb force of
 the point proton, (E) the Coulomb force of the proton with electric 
 form factor plus the vacuum polarization effect or (EM) in addition 
 to the interactions of the case E other tiny effects of the 
 electromagnetic interactions such as the interaction between the 
 anomalous magnetic moments of protons are involved, respectively.
 
 The relation between the effective range function $X_{0}(\nu)$ and 
 the phase shift of Eq.(15) must be modified.  Its explicit form 
 is given in other paper.\cite{present}\cite{heller}  
 What is important is that the effective 
 range function $X_{0}(\nu)$ is regular at $\nu=0$.  
 If we notice 
 this property, it is straightforward to construct 
 the once subtracted Kantor amplitude $K_{0}^{once}(\nu)$ which is 
 free from the singularity at $\nu=0$ when the interactions are  
 short range plus electromagnetic.\cite{present}  
 In order to ease the search of 
 the extra singularity, it is necessary to subtract the term of 
 the one-pion exchange at least approximately.
  The S-wave phase shift data $\delta_{0}^{EM}(\nu)$ of the energy 
 dependent analysis by Nijmegen group\cite{nij} 
 are used in the evaluation of 
 the once subtracted Kantor amplitude $K_{0}^{once}(\nu)$.   
 In figure 3,   $ - \tilde{K}_{0}^{once}(\nu) \equiv 
 -(K_{0}^{once}(\nu)-a_{0}^{once,1\pi}(\nu))$ is 
 plotted against $T_{lab}$, in which the range of $T_{lab}$ is 
 $0<T_{lab}< 125 \mbox{MeV.}$.  Figure 4 is the same graph with 
 enlarged scale and the narrow energy range $0<T_{lab}< 
 20 \mbox{MeV.}$.   In figure 4, the error bars are attached to 
 the data points (circles) of very low energy region, however for 
 higher energy the error becomes smaller than the size of the 
 data point.  We can clearly see a cusp at $\nu=0$ pointing upward, 
 which means the attractive long range force is acting.

 \begin{figure}[htp]
\begin{minipage}{6.8cm}
\includegraphics[width=.99\textwidth,height=5.0cm]{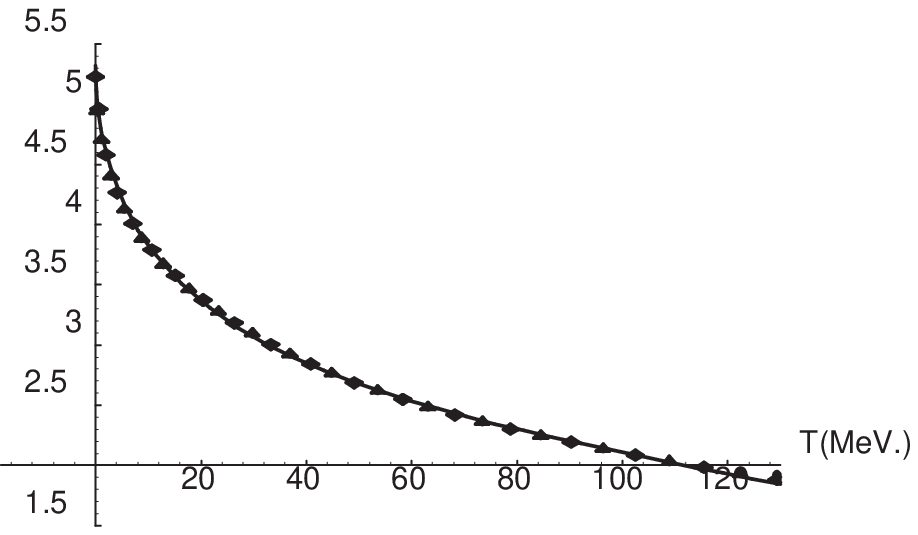}
\vskip-28mm
\hskip17mm
{\small - $\tilde{K}_{0}^{once}(\nu)$}

\vskip22mm

\caption{{\footnotesize
 $-\tilde{K}_{0}^{once}(\nu)$ is plotted against $T_{lab}$ in 
  $T_{lab} < 125 \; MeV.$, in which the Nijmegen phase shift 
 data of the energy dependent analysis are used.   
 The curve is the fit by the spectrum of the long 
range force in the range $ 0.6\;MeV. < T_{lab} <125\; MeV.$. 
The graph indicates the square root cusp at $T_{lab}=0$, which 
points upward. }}
\end{minipage}
\hfill
\begin{minipage}{6.8cm}
\includegraphics[width=.99\textwidth,height=5.0cm]{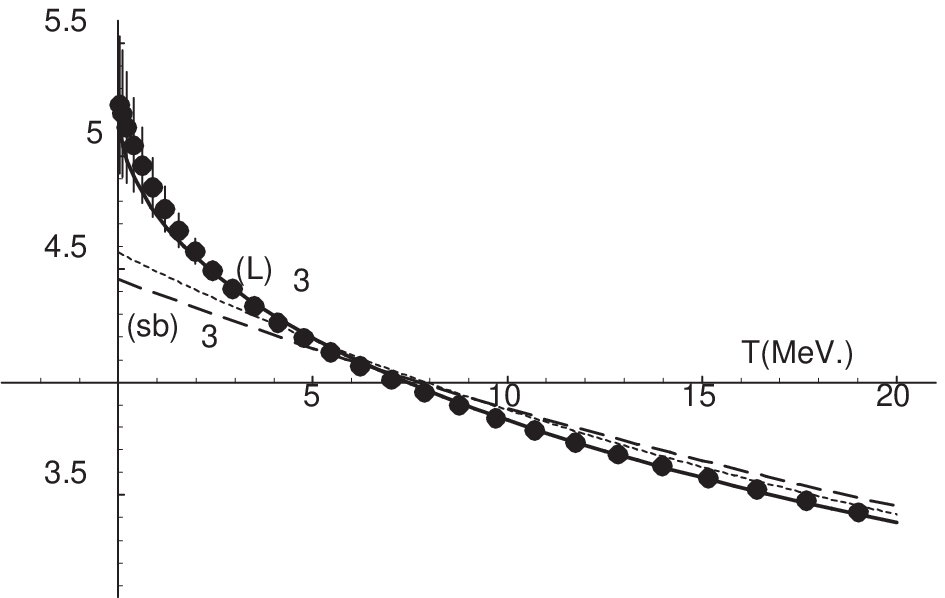}
\vskip-48mm
\hskip24mm
{\small - $\tilde{K}_{0}^{once}(\nu)$ }

\vskip5.5mm
\hskip16.0mm
\footnotesize $(sa)_{3}$

\vskip0.5mm
\hskip10.5mm
{ \large $\swarrow$ }

\vskip28.5mm

\caption{{\footnotesize
The enlarged graph of $-\tilde{K}_{0}^{once}(\nu)$  is plotted 
 in $ T_{lab} < 20 \; MeV.$.  The three curves are 
3-parameter fits to the data in $ 0.6\;MeV. < T_{lab} <125\;
 MeV.$. 
The curve $(L)_3$ is the fit by the spectrum of the long 
range force, whereas the curves of $(s \; a)_3$ (dots)
 and $(s \; b)_3$ (dash) 
are the fits by the short range potentials. }}
\end{minipage}
\end{figure}

 In order to determine the type of the long range force, let us 
 make the chi-square fit by the spectrum of the long range force 
 \begin{equation}
 A_{t}(s,t)= \pi C' t^{\gamma} e^{- \beta t}
 \end{equation}  
 Best fit in $0.6 \mbox{MeV.} <T_{lab}< 125 \mbox{MeV.}$ is achieved by   
 \begin{equation}
 \gamma=1.54 \quad , \quad C'=0.18 \quad and \quad \beta=0.063 
 \quad ,
 \end{equation}  
 in the unit of the neutral pion mass.   The curve of the best 
 fit is shown in fig.3 and also $(L)_{3}$ curve of fig.4, which 
 designates the long range spectrum with three parameters.   
 On the other hand, the fits by the short range force, whose 
 spectrum is the sum of three delta functions 
 and their coefficients are free parameters, are shown in fig.4. 
 For the dot curve $(sa)_{3}$, the locations of the delta functions 
 are $t=4, 9$ and $16$, whereas for the dash curve $(sb)_{3}$, the 
 locations are $t=9, 16$ and $25$ respectively.  It is evident that 
 the short range force cannot reproduce the data points well.

\section {Conclusion and prospect}
 
      Since the parameters of the spectrum of the long range force 
 are determined in Eq.(21), by using the relation Eq.(9) we 
 can evaluate the parameters of the tail of the long range the 
 potential $V(R) \sim -C/R^{\alpha}$, which become $\alpha=6.09$ and 
$C=0.18$ in the unit of the neutral pion mass.    
It indicates that reasonably strong Van der 
 Waals force of the London type is acting between nucleons. 
Therefore contrary to the ordinary model of the nuclear potential, 
in which only the short range forces are involved, 
it is important to construct nuclear potential anew 
based on the strong Van der Waals interaction plus the short range 
potential arising from the exchanges of mesons.

    If we remember the fact that the Van der Waals interaction is 
universal, the extra singularity at $t=0$ must occur in every 
amplitudes of the hadron-hadron reactions.  We can expect to 
observe it whenever very accurate data are available, or when 
we can construct a back ground function with wider domain of 
analyticity.  The low energy p-p amplitude is the former case,      
whereas the low energy P-wave amplitude of the pion-pion scattering  
belongs to the latter case.   Because of the crossing symmetry we 
can remove the cut of the two-pion exchange as well as the unitarity 
cut from the $\pi$-$\pi$ amplitude, therefore we can expect to 
observe the cusp in $a_{1}(\nu)/\nu$
 when reasonably accurate pion-pion 
 phase shifts data are given.  Details are found in separate 
 papers.\cite{kpipipb} 
   
      Finally from the upper bound of the strength $C$ of the 
Van der Waals potential given in Eq.(7), we can estimate the lower 
bound of the strength $^{*}e^2$ of the fundamental Coulomb 
interaction, it becomes 
\begin{equation}
   ^{*}e^2 > 3.06 \qquad , 
\end{equation}  
 in which the values of the nucleon radius $\bar{r}=0.5$ and of the 
 first excitation energy $\Delta E_{1}= m_{\Delta}-m_{N}=2.17$ are 
 used as well as the strength $C=0.18$ of the Van der Waals potential. 
 It is interesting to compare the result Eq.(22) with the strength 
 of the QCD, which is $^{*}e^2 \sim 0.3$, and also with that of the 
 magnetic monopole model of hadron, which is $^{*}e^2=137.0/4$ by 
 Dirac's charge quantization condition.\cite{dyon}  
 Therefore the value of the strength $C$ 
 of the Van der Waals potntial supports the magnetic monopole model of 
 hadron rather than the QCD.
 My dream is to confirm the magnetic momopole model of hadron\cite{dyon} 
by observing directly the 
monopoles of opposite  sign fuse to form a meson.

\end{document}